\documentclass[prb,onecolumn,showpacs,preprintnumbers,amsmath,amssymb]{revtex4}

\usepackage{graphicx}

\begin{document}

\title{Diffusion quantum Monte Carlo study of three-dimensional Wigner
crystals}

\author{N.~D.~Drummond, Z.~Radnai, J.~R.~Trail, M.~D.~Towler, and
R.~J.~Needs}

\affiliation{TCM Group, Cavendish Laboratory, University of Cambridge,
Madingley Road, Cambridge, CB3 0HE, UK}

\date{\today}

\begin{abstract}
We report diffusion quantum Monte Carlo calculations of
three-dimensional Wigner crystals in the density range $r_s=100$ to
$150$.  We have tested different types of orbital for use in the
approximate wave functions but none improve upon the simple Gaussian
form. The Gaussian exponents are optimized by directly minimizing the
diffusion quantum Monte Carlo energy.  We have carefully investigated and
sought to minimize the potential biases in our Monte Carlo results.
We conclude that the uniform electron gas undergoes
a transition from a ferromagnetic fluid to a body-centered cubic
Wigner crystal at $r_s=106 \pm 1$.  The diffusion quantum Monte Carlo
results are compared with those from Hartree-Fock and Hartree theory
in order to understand the role played by exchange and correlation in
Wigner crystals.  We also study ``floating'' Wigner crystals and give
results for their pair correlation functions.
\end{abstract}

 \maketitle

\section{Introduction}
\label{section:introduction}

Electrons in a uniform potential are expected to crystallize at low
densities to minimize their interaction
energy.\cite{wigner_1934,wigner_1938} In this paper we investigate
Wigner crystals in three dimensions using quantum Monte Carlo (QMC)
methods within the variational (VMC) and diffusion (DMC) approaches.
The DMC method is currently the most accurate available for
calculating the zero-temperature ground state energy of extended
quantum mechanical systems.

There has been some debate about the density at which the transition
from a Fermi fluid to a Wigner crystal should occur in three
dimensions.  In their pioneering DMC study of the phases of the
electron gas, Ceperley and Alder~\cite{ceperley_1980} obtained a
transition density of $r_s=100 \pm 20$,~\cite{units} but a more recent
DMC study gave $r_s=65 \pm 10$.\cite{ortiz_zero_temp_heg} Moreover,
highly accurate DMC energies for the low-density fluid have recently
become available~\cite{zong_2002}, which may further modify
predictions of the transition density. The primary goal of this work
is to provide highly accurate DMC energies for three-dimensional
Wigner crystals and to use them in conjunction with the fluid data of
Zong \textit{et al}.~\cite{zong_2002} to predict an accurate value for
the transition density.

To achieve sufficient accuracy we have carefully studied the possible
sources of error in our calculations, including finite size effects,
timestep errors and population control errors.  We have used three
procedures for optimizing the trial wave functions: minimization of
the variance of the energy within VMC~\cite{umrigar_1988,kent_1999};
minimization of the VMC energy; and minimization of the DMC energy.

We compare our results with recently-published fully-self-consistent
unrestricted Hartree-Fock (HF) calculations~\cite{trail_wigner}, and
with the results of a simple version of Hartree theory, which allows
us to understand the effects of exchange and correlation in Wigner
crystals.

Finally, we discuss ``floating'' Wigner crystals in which the
homogeneous and isotropic nature of the ground state is restored,
relating their properties to those of a ``fixed'' crystal.  DMC
results for the pair correlation functions (PCFs) of floating Wigner
crystals are compared with those of the fluid phases.

\section{Quantum Monte Carlo methods}

We have performed both VMC and DMC calculations using the CASINO
package.\cite{casino} In VMC, expectation values are calculated using
an approximate many-body wave function, the integrals being performed
by a Monte Carlo method.  The approximate wave function normally
contains a number of variable parameters, whose values are obtained by
an optimization procedure.

In the DMC method~\cite{ceperley_1980,foulkes_2001} the imaginary time
Schr\"odinger equation is used to evolve an ensemble of electronic
configurations towards the ground state.  The
fermionic symmetry is maintained by the fixed-node
approximation~\cite{anderson_1976}, in which the nodal surface of the
wave function is constrained to equal that of an approximate wave
function.  We will refer to the approximate wave functions used in VMC
and DMC as trial wave functions.  The fixed-node DMC energy provides a
variational upper bound on the ground state energy with an error that
is second order in the error in the nodal
surface.\cite{moskowitz_1982,reynolds_1982}

The trial wave function introduces importance sampling and controls
both the statistical efficiency and the final accuracy that can be
obtained. In DMC the final accuracy depends on the nodal surface of
the trial wave function via the fixed-node approximation, while in VMC
the final accuracy depends on the entire trial wave function, so that
VMC energies are more sensitive to the quality of the trial wave
function than DMC energies.  Apart from the fixed-node approximation,
DMC results may be subject to bias from the use of the short-time
approximation (finite timestep errors), population control errors, and
effects from the finite size of the simulation cell.  We have made
strenuous attempts to reduce these errors: see
Sec.~\ref{section:accuracy}.  The statistical errors in our QMC data
are estimated using the blocking method~\cite{flyvbjerg_1989} to
eliminate the effects of serial correlation.

\section{Trial wave functions}

\subsection{The Slater-Jastrow form}

We have used trial wave functions of the standard Slater-Jastrow form,
\begin{equation}
\label{eq:slater-jatrsow}
\Psi \left( {\bf R}_\uparrow, {\bf R}_\downarrow \right)=e^{J\left(
{\bf R}_\uparrow, {\bf R}_\downarrow \right)} D_\uparrow \left( {\bf
R}_\uparrow \right) D_\downarrow \left( {\bf R}_\downarrow \right)\ ,
\end{equation}
where $D_\uparrow$ and $D_\downarrow$ are Slater determinants of up
and down spin single-particle orbitals and ${\bf R}_\uparrow$ and
${\bf R}_\downarrow$ denote the coordinates of the up and down spin
electrons, respectively, and $e^{J}$ is the Jastrow correlation
factor.

\subsection{Jastrow factors \label{subsec:jastrow}}

We have used Jastrow factors of the form
\begin{equation} 
J=-\frac{1}{2} \sum_i \sum_j \left( u_0(r_{ij}) + S_1(r_{ij}) \right),
\end{equation}
\noindent where
\begin{equation} 
\label{eq:u0}
u_0(r_{ij})=\frac{A}{r_{ij}} \left( 1 - \exp
\left(-\frac{r_{ij}}{F_{ij}} \right) \right) \exp \left(
-\frac{r_{ij}^2}{L_0^2} \right),
\end{equation}
\noindent with $F_{ij}=\sqrt{2A}$ if electrons $i$ and $j$ have the
same spin and $F_{ij}=\sqrt{A}$ if the electrons have opposite
spins. This term satisfies the electron-electron cusp
conditions.\cite{kato_1957,foulkes_2001} The constant $L_0$ is set to
$0.3$ of the Wigner-Seitz radius of the simulation cell and $A$ is a
free parameter. The $u_0$ term is set to zero for $r_{ij}$ greater
than the Wigner-Seitz radius, resulting in a small discontinuity in
the Jastrow factor of less than $2 \times 10^{-5}$ in magnitude.  To
investigate possible bias from this discontinuity we compared the
values of the two standard estimators of the kinetic energy involving
the gradient and Laplacian of the trial wave function from a very long
VMC run at $r_s=100$.  The estimators agreed to within the statistical
error, which was smaller than in our final DMC runs.

The second term in the Jastrow factor is given by
\begin{eqnarray} 
S_1(r_{ij}) & = & (r_{ij}-L^\prime)^2 r_{ij}^2 \sum_{l=0}^{L} \alpha_l
T_l \left( \frac{2r_{ij}-L^\prime}{L^\prime} \right) \nonumber \\ & &
+ B^\prime (r_{ij}-L^\prime)^2 \left( \frac{L^\prime}{2}+r_{ij}
\right),
\end{eqnarray}
\noindent where $T_l$ is the $l$th Chebyshev polynomial, $L^\prime$ is
the Wigner-Seitz radius of the simulation cell and $B^\prime$ and the
$\alpha_l$ are parameters to be determined.

\subsection{Orbitals for the Slater determinants}

The Slater determinants for the crystalline phases were formed from
localized non-orthogonal single-particle orbitals centered on the
lattice sites of a body-centered cubic (BCC) crystal. A BCC crystal is
expected to be favored in the low-density limit because it has the
lowest Madelung energy.  Throughout, we use $\phi({\bf r})$ to denote
a spatial orbital centered on the origin. Periodic orbitals for use in
a simulation of a finite system subject to periodic boundary
conditions are constructed for each lattice point in the simulation
cell by summing over all the replicas of $\phi$ centered on that
lattice point. Clearly, if all the individual orbitals are periodic
then their Slater determinant is too.  We use a Jastrow factor
containing only homogeneous terms, see Sec.~\ref{subsec:jastrow}, with
the differences between electron coordinates being evaluated under the
minimum image convention. Hence the overall wave function is periodic.

In their VMC and DMC studies of Wigner crystals,
Ceperley~\cite{ceperley_fermion_plasma} and Ceperley and
Alder~\cite{ceperley_1980} used Gaussian orbitals:
\begin{equation} 
\label{eq:gaussian}
\phi({\bf r}) = \exp(-Cr^2).
\end{equation}
They determined $C$ by variational methods, with the Jastrow factor
being optimized simultaneously.

Ortiz \textit{et al}.~\cite{ortiz_zero_temp_heg} used exponentials of
two-parameter Pad\'e functions:
\begin{equation} 
\label{eq:pade}
\phi({\bf r}) = \exp \left( \frac{-C r^2}{1+kr} \right),
\end{equation}
\noindent and determined the values of $C$ and $k$ and the parameters
in their Jastrow factor by minimizing the variance of the energy
within VMC. 

We have also investigated two new types of orbital for Wigner
crystals. A straightforward generalization of Eq.~(\ref{eq:gaussian})
is to use a linear combination of Gaussian orbitals:
\begin{equation} 
\phi({\bf r}) = \sum_{i=1}^{N_G} \lambda_i \exp \left( -C_i r^2
\right).
\label{eqn:multigauss_hf} 
\end{equation} 
We have also considered orbitals based on an expansion in the
eigenstates of a simple harmonic oscillator.  An orbital is
constructed by multiplying the simple Gaussian function by a
polynomial.  For a BCC lattice with identical orbitals on each site
the polynomial should have the full symmetry of the lattice, i.e.,
\begin{eqnarray} 
\phi({\bf r}) = \exp(-Cr^2) \big[ 1 + \alpha r^2 + \beta r^4 + \gamma
(x^2y^2+x^2z^2+y^2z^2) + {\cal{O}}(r^6) \big],
\label{eqn:harmonic} 
\end{eqnarray}
\noindent where ${\bf r}=(x,y,z)$.  This orbital has considerable
flexibility at small $r$.

\subsection{Optimization of the trial wave functions \label{subsec:orb_opt}}

Parameters in the trial wave functions may be optimized in a variety
of ways.  In principle the DMC energy depends only on the nodal
surface of the wave function, which is determined by the form of the
orbital.  It is therefore best to minimize the DMC energy directly
with respect to the parameters in the orbitals, but this is a costly
and laborious procedure which we have carried out only for the
Gaussian parameter $C$ of the simple Gaussian orbital.  In principle
the DMC energy does not depend on the Jastrow factor, so it cannot be
optimized in this fashion.

We first studied the simple Gaussian and Pad\'e forms of
Eqs.~(\ref{eq:gaussian}) and (\ref{eq:pade}), using energy variance
minimization to optimize the $C$ parameter and the parameters in
Jastrow factor, and minimization of the VMC energy with respect to
variations in $k$, but we found the optimal value of $k$ to be very
close to zero.  The Pad\'e orbital seems to offer little advantage
over a simple Gaussian orbital at the densities studied ($100 \leq r_s
\leq 150$).

We optimized the expansion coefficients and Gaussian exponents of the
linear combination of Gaussian orbitals form of
Eq.~(\ref{eqn:multigauss_hf}), together with a Jastrow factor, using
VMC energy variance minimization. However, again, it was found that in
practice this orbital offers no advantage over a single Gaussian
function at the densities studied.

We attempted to optimize the $\alpha$ parameter in the harmonic
oscillator form of Eq.~(\ref{eqn:harmonic}) for a Wigner crystal at
$r_s=100$, but the resulting wave function gave a DMC energy within
the error bars of the one obtained by setting $\alpha=0$.


We have therefore adopted the simple Gaussian orbital of
Eq.~(\ref{eq:gaussian}) for our main calculations.  We have adopted
the following procedure to optimize the trial wave functions.  The $A$
parameter in the Jastrow factor was optimized by minimization of the
VMC energy, the parameters in the $S_1$ part of the Jastrow factor by
energy variance minimization, and the $C$ parameter in the Gaussian
orbitals by minimization of the DMC energy.  These minimizations were
performed in turn until the changes in the parameters were negligible.
We found that the DMC-optimized exponents obey $C \approx 0.11
r_s^{-3/2}$, which gives rather smaller values than those used by
Ceperley~\cite{ceperley_fermion_plasma,ceperley_1980} ($C \approx 0.2
r_s^{-3/2}$) and considerably smaller than those predicted by HF
theory or the simple Hartree model ($C \approx 0.5 r_s^{-3/2}$), see
Sec.~\ref{section:hf}.

\section{Accuracy of the DMC results \label{section:accuracy}}

Our DMC algorithm is essentially that of Umrigar \textit{et
al}.\cite{umrigar_1993} Here we explore the sources of error in our
DMC calculations and justify our choices of the parameters for the
final production runs, which are summarized in
Sec.~\ref{subsection:parameters}. Unless otherwise stated, we use
simple Gaussian orbitals throughout this section.

\subsection{Finite size effects
\label{subsection:finite_size}}

We used periodic boundary conditions and the Ewald interaction energy
to reduce the finite size effects.  We tested the convergence of the
Ewald sums and found that truncation errors were less than $10^{-3}$
of the statistical error bars on the final DMC runs.

The energy per electron at a given density depends on the number of
electrons in the simulation cell. We wish to obtain the energy per
electron in the limit that the number of electrons per simulation cell
goes to infinity.  Two approaches have been used previously when
dealing with finite size effects in QMC simulations of Wigner
crystals. Ortiz \textit{et al}.~\cite{ortiz_zero_temp_heg} used large
simulation cells and found the finite size errors to be less than
their statistical error bars of $8.5 \times 10^{-6}$\,a.u.~per
electron for numbers of electrons in excess of 500.
Ceperley~\cite{ceperley_fermion_plasma,ceperley_1980}, on the other
hand, used smaller system sizes in conjunction with an extrapolation
scheme.

Because we wish to work to very high accuracy, we use quite large
simulations cells and the extrapolation formula derived by
Ceperley~\cite{ceperley_fermion_plasma}:
\begin{equation} 
E_\infty = E_N + \frac{b}{r_s^{3/2}N},
\label{eqn:cep_extrap_formula} 
\end{equation}
\noindent where the constants $E_\infty$ and $E_N$ are the total
energies per electron of the infinite system and the system with $N$
electrons, and $b$ is independent of both $r_s$ and $N$.

Starting from a two-parameter Pad\'e orbital and corresponding Jastrow
factor optimized in a 64-electron simulation cell, we attempted to
further optimize the wave function in a 216-electron unit cell. This
attempt did not lead to a lowering of the VMC energy, suggesting that
the 64-electron simulation cell is adequate for optimization purposes;
this size of cell was used in all of our subsequent optimization runs.

\subsection{Population control biasing \label{section:pop_control_bias_wc}}

\begin{figure}[ht]
\includegraphics[width=10cm]{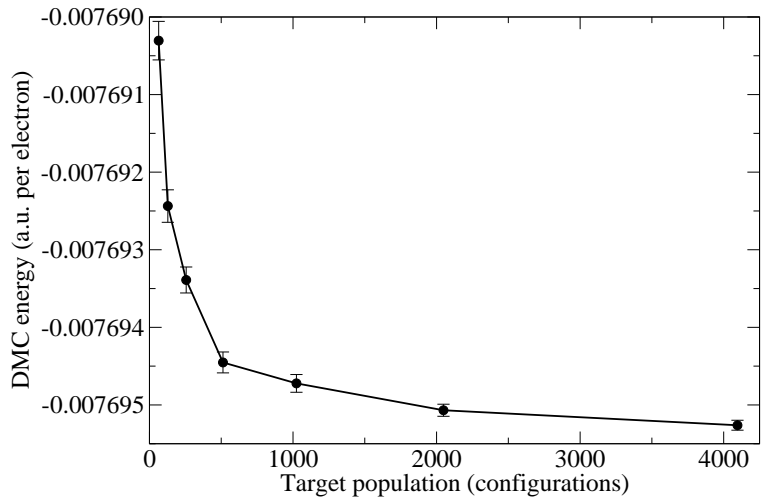}
\caption{DMC energy against target population for a 64-electron
crystal at $r_s=100$. The Gaussian exponent was $C=0.000135$ and the
Jastrow factor contained only the $u_0$ term, with $A=438.389$. The
timestep was $20$\,a.u.
\label{graph:population_control_bias}}
\end{figure}

The use of a finite population of configurations results in a positive
bias in the DMC energy which, it is argued, falls off as the
reciprocal of the target population.\cite{umrigar_1993} This turns out
to be a genuine problem in the case of Wigner crystals where we are
able to work to extremely high precision.  An example of the problem
of population control biasing is shown in
Fig.~\ref{graph:population_control_bias}, where it can be seen that
the bias is indeed positive and that it falls off roughly as the
reciprocal of the target population.

The simplest method for avoiding population control biasing is to use
a large target population. Alternatively, the reweighting scheme
developed by Umrigar \textit{et al}.~\cite{umrigar_1993} can be used.
In our tests we found this scheme to work well, provided the number of
reweighting factors was about the same as the number of timesteps over
which average local energies are correlated (between 100 and 1000
timesteps of $30$\,a.u.). However, with this many reweighting factors
present, the total weight at each timestep fluctuated enormously and
very long simulations were required in order to obtain meaningful
statistics. We found it to be more efficient to use larger populations
than to employ the reweighting scheme.  For this reason, we did not
use the reweighting scheme in our production runs.

\begin{figure}[ht]
\includegraphics[width=10cm]{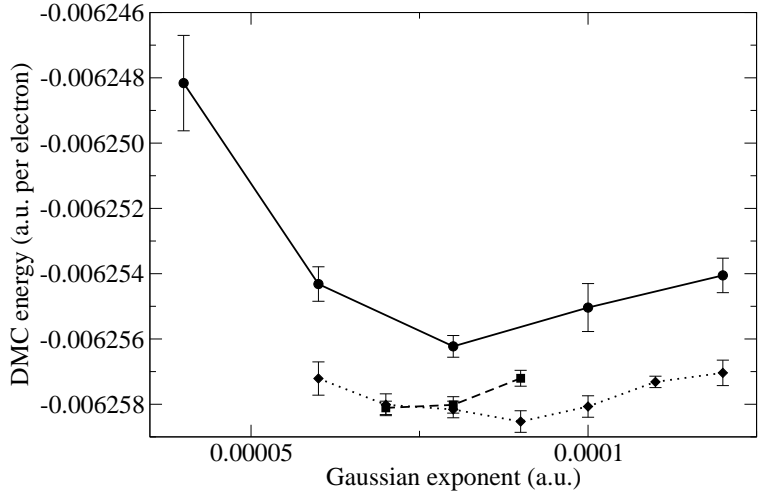}
\caption{DMC energy against Gaussian exponent for a 64-electron
crystal at $r_s=125$, using a timestep of 30\,a.u.~and $A=597.901$.
Solid line: DMC energies with a Jastrow factor consisting of only the
$u_0$ term and a target population of 100 configurations; dashed line:
DMC energies for the same Jastrow factor, but with a target population
of 800 configurations; dotted line: DMC energies obtained with a fully
optimized Jastrow factor (with both $u_0$ and $S_1$ terms) and a
target population of 100 configurations.
\label{graph:dmc_energy_v_exp_rs125_antiferro}}
\end{figure}

Including more parameters in the Jastrow factor can lead to a
significant reduction in population control biasing, as illustrated in
Fig.~\ref{graph:dmc_energy_v_exp_rs125_antiferro}. This shows that,
when the reweighting scheme is not used, DMC energies obtained with a
a poor Jastrow factor and a small target population (solid line) are
too high, but when a larger population is used in conjunction with the
same Jastrow factor (dashed line) the energies are similar to those
obtained with a good Jastrow factor and a small population (dotted
line).  Improving the overall quality of the trial wave function
reduces the population control bias because it reduces the
fluctuations in the population.  The results shown in
Fig.~\ref{graph:population_control_bias} in which the Jastrow factor
consisted only of the $u_0$ term therefore represent a worst-case
scenario.

\subsection{Timestep biasing}

\begin{figure}[ht]
\includegraphics[width=10cm]{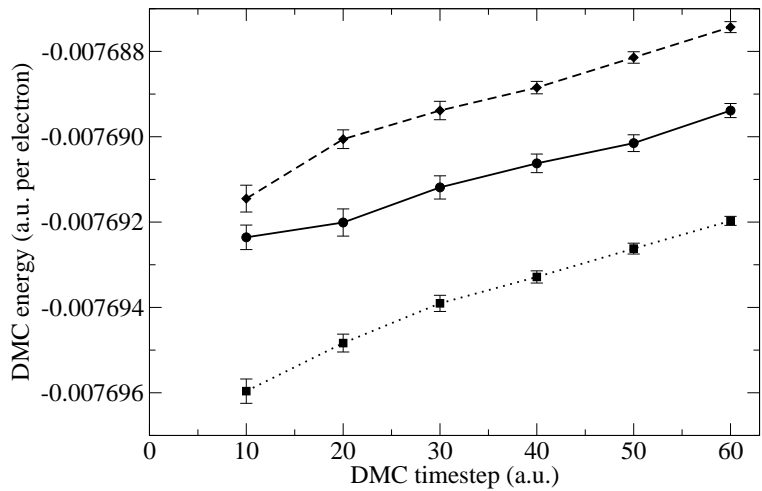}
\caption{DMC energy of a 64-electron crystal at $r_s=100$ against
timestep for different values of the $A$-parameter in the Jastrow
factor of Eq.~(\ref{eq:u0}) and different target population sizes. The
$S_1$ term was not present in the Jastrow factor. The Gaussian
exponent was $C=0.00011$ in all cases. Solid line: $A=438.389$, target
population 100 configurations; dotted line: $A=438.389$, target
population 800 configurations; dashed line: $A=563.157$, target
population 100 configurations.}
\label{graph:wigner_dmc_energy_v_tau}
\end{figure}

The variation of DMC energy with timestep is shown in
Fig.~\ref{graph:wigner_dmc_energy_v_tau} for three different Jastrow
factors.  It can be seen that the bias is always positive and that it
grows linearly with timestep; therefore we can largely eliminate the
bias by carrying out simulations at a number of different timesteps
and performing a linear extrapolation to zero timestep.

The differences between the DMC energies in
Fig.~\ref{graph:wigner_dmc_energy_v_tau} are due to population control
and timestep bias.  The solid and dashed curves were obtained using
the same target population of 100 configurations, but with different
Jastrow factors.  The population control bias at fixed timestep is
clearly smaller for the Jastrow factor with $A=438.389$ (solid line)
than for $A=563.157$ (dashed line).  The dotted line was obtained with
a target population of 800 configurations, which essentially removes
the population control bias.  Because the solid and dotted curves are
approximately parallel we deduce that the population and timestep
errors are approximately independent of one another.  Furthermore, it
is clear that altering the Jastrow factor has a considerably greater
effect on the population control bias than on the timestep bias.

\begin{figure}[ht]
\includegraphics[width=10cm]{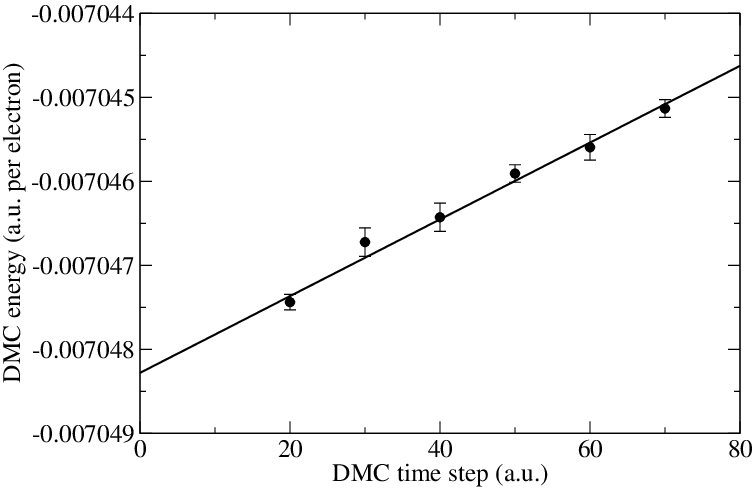}
\caption{DMC energy of a 64-electron crystal at $r_s=110$ against
timestep. The Jastrow factor contained optimized $u_0$ and $S_1$ terms
and the target population was 640 configurations. The straight line is
a fit to the DMC data.}
\label{graph:energy_v_tau_AF_110}
\end{figure}

In Fig.~\ref{graph:energy_v_tau_AF_110} we show that timestep bias
remains a problem even with a well-optimized Jastrow factor and a
large target population which essentially removes the population
bias. However, this figure again shows that a linear fit is
appropriate when extrapolating to zero timestep.

\subsection{Parameters for the production runs \label{subsection:parameters}}

The final production runs were characterized as follows.

\begin{enumerate}

\item The Gaussian exponents in the orbitals were optimized by
  minimizing the DMC energy, the $A$ parameters by minimizing the VMC
  energy, and the other parameters in the Jastrow factors by
  minimizing the variance of the energy, as described in
  Sec.~\ref{subsec:orb_opt}.  The $S_1$ terms in the Jastrow factors
  contained between 4 and 6 parameters per spin.

\item A target population of 640 configurations was used. This,
  together with the optimized Jastrow factor, should ensure that
  population control errors are negligible.

\item At each density, DMC calculations were performed using between
  four and six different timesteps and the energy was extrapolated
  linearly to zero timestep.

\item A variety of system sizes were used (see
  Table~\ref{table:zero_tau_dmc_energies}), and the energies were
  extrapolated to infinite system size using
  Eq.~(\ref{eqn:cep_extrap_formula}).

\end{enumerate}

\section{Results and discussion \label{section:results}}

\subsection{DMC energies obtained using DMC-optimized Gaussian orbitals
\label{section:dmc_results}}

The values of the exponents obtained by optimizing the DMC energy are
shown in Table~\ref{table:opt_exponents_and_dmc_energies}, along with
the final DMC energies. These were found by using the DMC results
shown in Table~\ref{table:zero_tau_dmc_energies} in conjunction with
the finite size extrapolation formula of
Eq.~(\ref{eqn:cep_extrap_formula}). Using the results for $r_s=100$
and $r_s=125$, we obtained a good fit, giving $b=1.26(3)$.

\begin{table}[ht]
\begin{tabular}{rrr}\hline\hline
$r_s$ & $C$ (DMC) & DMC energy \\ \hline

100 & $0.00011$ & $-0.0076765(4)$ \\

110 & $0.0001$ & $-0.0070312(5)$ \\

125 & $0.00009$ & $-0.0062458(4)$ \\

150 & $0.000063$ & $-0.0052690(3)$\\ \hline\hline
\end{tabular}
\caption{Orbital exponents optimized by minimizing the DMC energy, and
the final DMC energies per electron (extrapolated to zero timestep and
infinite system size) for the different $r_s$. All entries are in
a.u. \label{table:opt_exponents_and_dmc_energies}}
\end{table}

\begin{table}[ht]
\begin{tabular}{rrr}\hline\hline
$r_s$ & System size & DMC energy \\ \hline

100 & 64 & $-0.0076961(2)$ \\

100 & 216 & $-0.0076823(3)$ \\

100 & 512 & $-0.0076788(8)$ \\

110 & 64 & $-0.0070483(2)$ \\

125 & 64 & $-0.0062599(2)$ \\

125 & 216 & $-0.0062495(1)$ \\

150 & 64 & $-0.0052797(1)$ \\ \hline\hline

\end{tabular}
\caption{DMC energies in a.u. per electron (extrapolated to zero
timestep) at different densities and system sizes, which are
characterized by the number of electrons in the simulation cell.
\label{table:zero_tau_dmc_energies}}
\end{table}

\subsection{Electronic charge densities obtained using the different
  orbitals \label{section:charge_densities}}

\begin{figure}[ht]
\includegraphics[width=10cm]{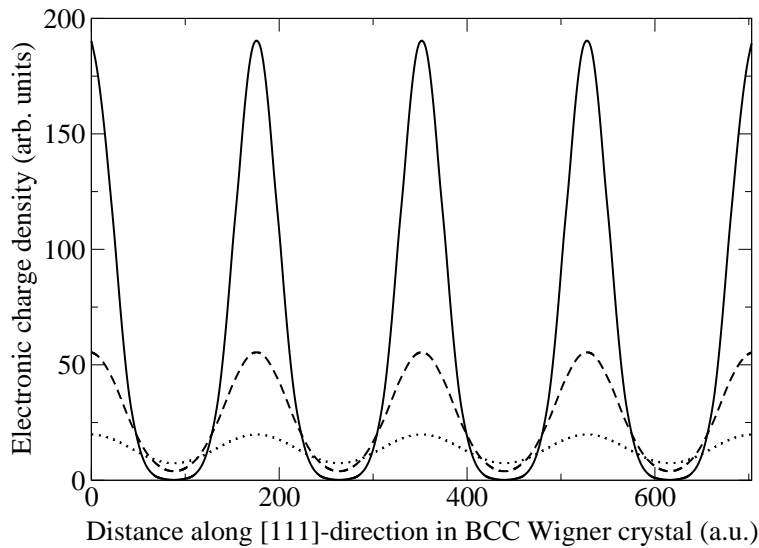}
\caption{Electronic charge density along a $\left<111\right>$
direction passing through the lattice sites of a BCC Wigner crystal at
$r_s=100$. Solid line: HF charge density; dotted line: charge density
from the DMC-optimized orbitals without a Jastrow factor; dashed line:
DMC charge density.
\label{graph:linedensity_wigner_rs100_ferro}}
\end{figure}

In Fig.~\ref{graph:linedensity_wigner_rs100_ferro} we plot the
electronic charge densities for a $r_s=100$ crystal, calculated using
HF theory orbitals (see Sec.~\ref{section:hf}), from the DMC-optimized
orbitals but without a Jastrow factor, and within DMC using an
optimized Jastrow factor.

The HF theory orbitals are very localized whereas the orbitals
optimized within DMC are much more diffuse. However, the inclusion of
a Jastrow factor results in the charge density from the DMC-optimized
orbitals becoming more localized on lattice sites, although not to the
same extent as the HF orbitals.  The VMC charge density obtained with
the optimized Slater-Jastrow wave function is very close to the DMC
density shown in Fig.~\ref{graph:linedensity_wigner_rs100_ferro}.  The
peak difference between the DMC and VMC charge densities is 2.6
arb. units, and therefore an extrapolated estimation of the QMC charge
density, $\rho(r) \approx 2 \rho_{\rm{DMC}} - \rho_{\rm{VMC}}$, would
be very similar to the DMC density.

The Jastrow factor in a Wigner crystal serves to further localize the
electrons on their lattice sites. Therefore, whereas HF theory gives
very localized Gaussian orbitals, in an optimized Slater-Jastrow wave
function we find the \textit{orbitals} to be much more diffuse, with
the localization being caused by correlation effects from the Jastrow
factor.  Correlation effects allow electrons to invade each other's
space to some extent without incurring a high potential energy
penalty; hence the overall charge density is less localized than in HF
theory, and the kinetic energy is lower as a result.

\section{Locating the fluid-to-crystal transition density
\label{section:fluid_crystal_fitting_functions}}

In order to locate the density at which the transition from the Fermi
fluid to the crystal occurs, we fit the DMC energy data for these
phases to interpolating functions.  The correlation energy is defined
as
\begin{equation} 
E_c(r_s,\zeta)=E(r_s,\zeta)-E_{HF}(r_s,\zeta),
\end{equation}
\noindent where $E$ is the total energy, $E_{HF}$ is the HF ground
state energy, $\zeta$ is the polarization.
Ceperley~\cite{ceperley_fermion_plasma} proposed a fitting form for
the correlation energy of a Fermi fluid as a function of $r_s$,
\begin{equation}
E_c^\zeta(r_s)=\frac{\gamma_\zeta}{1+\beta_1^\zeta\sqrt{r_s}+\beta_2^\zeta
r_s},
\label{eqn:cep_fit_fluid} 
\end{equation}
\noindent where $\gamma_\zeta$, $\beta_1^\zeta$ and $\beta_2^\zeta$
are fitting parameters.  We make use of the highly accurate DMC
energies of Zong \textit{et al.}~\cite{zong_2002} for the
ferromagnetic fluid phase at low densities, which used trial wave
functions including ``backflow'' effects.  We found an excellent fit
to Eq.~(\ref{eqn:cep_fit_fluid}) giving $\gamma_1=-0.09399$,
$\beta_1^1=1.5268$ and $\beta_2^1=0.28882$. We also tried fitting the
fluid data to the form proposed by Perdew and Zunger \cite{perdew},
which is based upon Eq.~(\ref{eqn:cep_fit_fluid}), but includes an
assumed dependence on polarization, so that the partially polarized
and unpolarized data of Zong \textit{et al}.~could be used. The
$\chi^2$ value of this fit was not so good, however.

At low densities the total energy of a crystal phase can be expanded as
\begin{equation}
E(r_s) =
\frac{f_0}{r_s}+\frac{f_1}{r_s^{3/2}}+\frac{f_2}{r_s^2}+{\cal{O}}(r_s^{-5/2}),
\label{eqn:low_density_exp} 
\end{equation}
\noindent where the $\{ f \}$ are constants.\cite{carr_crystal} The
first of these is taken to be $f_0=-0.89593$ in order to give the
Madelung energy in the low density limit.  We found that our DMC data
fitted Eq.~(\ref{eqn:low_density_exp}) very well, giving $f_1=1.3379$
and $f_2=-0.55270$.  These values are in reasonable agreement with the
parameters found using a completely different method by
Carr~\cite{carr_crystal} and Carr, Coldwell-Horsfall and
Fein~\cite{carr_anharmonic}, who have calculated the zero-point
lattice-vibrational energy of a Wigner crystal in order to give an
analytical result of $f_1=1.325$. Furthermore, they use perturbation
theory to obtain the approximate result $f_2=-0.365$.  This phonon
model is in good agreement with our Wigner crystal energies at large
$r_s$, but it gives energies which are too high at smaller $r_s$.

\begin{figure}[ht]
\includegraphics[width=10cm]{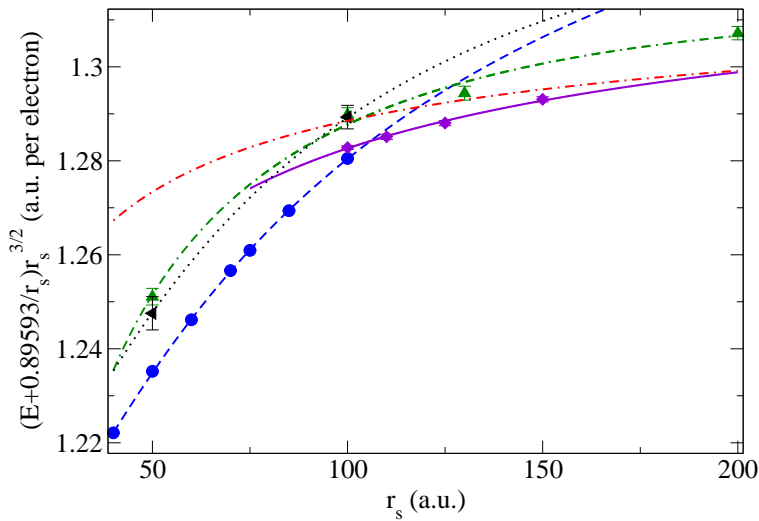}
\caption{Energies of the ferromagnetic fluid and BCC crystalline
phases at low density. The Madelung energy of the BCC lattice has been
subtracted off and the resulting energy multiplied by $r_s^{3/2}$ to
highlight the differences between phases.  The circles show the DMC
data of Zong \textit{et al}.\cite{zong_2002} for the ferromagnetic
Fermi fluid; the dashed line is a fit to this data.  The diamonds show
our DMC results for the BCC crystal; the solid line is a fit to this
data.  The left-pointing triangles show the
Ceperley-Alder~\cite{ceperley_1980} results for the ferromagnetic
fluid; the dotted line is a fit to this data.  The up-pointing
triangles show the Ceperley-Alder results for the BCC crystal; the
dashed-dashed-dotted line is a fit to this data.  Finally, the
dashed-dotted line shows the prediction of the phonon model of Carr
\textit{et al}.\cite{carr_anharmonic} Where error bars on the DMC data
cannot be seen, it is because they are smaller than the symbols
showing the data points.
\label{graph:fluid_to_crystal_transition}}
\end{figure}

The energies of the ferromagnetic fluid and BCC crystalline phases at
low densities calculated by different authors are shown in
Fig.~\ref{graph:fluid_to_crystal_transition}. We found the transition
from the fluid to crystalline phases to occur at $r_s=106 \pm 1$, in
agreement with the original result of Ceperley and
Alder.\cite{ceperley_1980} Note, however, that the transition density
predicted using the fluid data of Zong \textit{et
al}.~\cite{zong_2002} in conjunction with the crystal data of Ceperley
and Alder would be somewhat lower, at about $r_s=127$.  Our Wigner
crystal energies are slightly lower than those of Ceperley and Alder,
even though they studied a Bose crystal, which must have a lower
energy than the corresponding fermion one.  We believe this difference
must be due to some small bias in the results of Ceperley and
Alder.\cite{ceperley_1980} Fitting Eq.~(\ref{eqn:low_density_exp}) to
the crystal data of Ceperley and Alder, we find
that $f_1=1.4309$ and $f_2=-1.1058$. The discrepancy with the
analytical result of Carr for $f_1$ is consistent with the presence
of a small, positive, systematic bias in the crystal results of
Ceperley and Alder.

Our transition density is considerably lower than the value of
$r_s=65 \pm 10$ obtained by Ortiz \textit{et
al.}~\cite{ortiz_zero_temp_heg,harris_thesis}.  The statistical error
bars on their data are much larger than on the fluid data of
Zong \textit{et al}.~\cite{zong_2002} or on our Wigner crystal data,
which hampers detailed comparisons.  However, it appears that the main
reason for the discrepancy is that Ortiz \textit{et al.}~place the
ferromagnetic fluid higher in energy than Zong \textit{et al}.  Some
bias in the results of Ortiz \textit{et al.}~is expected because they
used plane wave nodes while Zong \textit{et al.}~used optimized
backflow nodes, but this is not enough to explain the difference
between the results.

For the crystal phase, we may estimate the fixed node error resulting
from the use of the HF orbitals by taking the difference between DMC
energies calculated using the HF and DMC-optimized orbitals. At
$r_s=100$ we find this difference to be $8.9(1)\times
10^{-5}$\,a.u.~per electron.  Zong \textit{et
al}.~\cite{zong_unpublished} have calculated the fixed node errors
resulting from the use of plane-wave orbitals for the fluid phases as
$1.87(5)\times 10^{-5}$\,a.u.~per electron for the unpolarized fluid
and $0.84(7)\times 10^{-5}$\,a.u.~per electron for the fully polarized
fluid.  Therefore, although the correlation energy of crystals is
smaller than fluids at the same density, the fixed node errors
resulting from the use of HF orbitals are considerably larger in
crystals than in fluids.

\section{HF and other simple theories \label{section:hf}}

HF theory is described as ``restricted'' when the spin-orbitals are
products of space and spin parts which are occupied in pairs with
identical space parts, and ``unrestricted'' when the space parts are
different or when they are not occupied in pairs.  Within the quantum
chemistry community the standard definition of the correlation energy
is the difference between the exact and restricted HF energies, but in
a Wigner crystal the electrons are localized in space individually and
a description within restricted HF theory is not possible.  We
therefore define the correlation energy as the difference between the
exact and unrestricted HF energies.

A recent self-consistent unrestricted HF study of electrons in a
uniform potential gave stable Wigner crystal solutions for $r_s \geq
4.5$ in three dimensions.\cite{trail_wigner} Here we develop
simplified versions of the HF model which almost exactly reproduce the
results of the fully self-consistent HF studies~\cite{trail_wigner} at
low densities and compare the results with our DMC ones.

In the low-density regime, the overlap between orbitals centered on
neighboring lattice sites is small and therefore we expect the Hartree
and HF energies to be similar. Let us take the wave function of the
crystalline phase to be a Hartree product of normalized Gaussian
orbitals,
\begin{equation} 
\phi({\bf r})=\left( \frac{2C}{\pi} \right)^{3/4} \exp( -C r^2),
\end{equation}
\noindent centered on lattice sites. The Gaussian exponent $C$ is to
be determined variationally.

The resulting kinetic energy per electron is easily evaluated as
$T=3C/2$.  The spatial charge density is simply the superposition of
the Gaussian charge densities due to each orbital. The electrostatic
energy $E_H$ of this charge distribution may be readily evaluated, but
we must subtract off the self-energies $E_0$ of each Gaussian. The
total energy per electron is therefore given by
\begin{eqnarray} 
E & = & T+E_H-E_0 \nonumber \\ & = & \frac{3 C}{2} + \frac{2
\pi}{\Omega} \sum_{{\bf G} \neq {\bf 0}} \frac{\exp(-G^2/4C)}{G^2} -
\sqrt{\frac{C}{\pi}}.
\label{eqn:simple_model} 
\end{eqnarray}
\noindent where the ${\bf G}$ are the reciprocal lattice vectors and
$\Omega=4 \pi r_s^3/3$ is the volume of the primitive unit cell.

Differentiating this energy with respect to the Gaussian exponent and
approximating the sum by an integral, we find that
\begin{equation} 
\frac{\partial E}{\partial C} \approx \frac{3}{2}-\frac{\pi}{2\Omega C^2}.
\label{eqn:dE_dC_simple} 
\end{equation}
\noindent The approximation is valid for large $r_s$, where the
density of reciprocal lattice vectors is large. Hence we find the
optimal value of
$C$ to be $C = 1 / 2 r_s^{3/2}$, which is precisely the result obtained by
Wigner~\cite{wigner_1938} using a spherical approximation.

Integrating Eq.~(\ref{eqn:dE_dC_simple}) with respect to $C$ we obtain
the energy $E = 3C/2+\pi / 2 \Omega C + f(r_s)$, where the function
$f(r_s)$ is the ``constant'' of integration. Inserting the optimal
value of $C$ and making use of the fact that, in the limit of low
densities, $E$ must tend to the Madelung energy of the crystal
lattice, we find
\begin{equation} 
E = \frac{3}{2 r_s^{3/2}}+\frac{M}{r_s},
\label{eqn:approx_simple_model} 
\end{equation}
\noindent where $M$ is the Madelung constant of the lattice.

The simple Hartree model of Eq.~(\ref{eqn:simple_model}) and the
further simplifying approximation of
Eq.~(\ref{eqn:approx_simple_model}) give energies which are very close
to the full HF results for $r_s>50$; for example, the energy of
Eq.~(\ref{eqn:simple_model}) agrees with that of the full HF result to
within 0.006\%, whereas the energies at $r_s=100$ agree to within
0.001\%.  The agreement between the fully self-consistent HF data and
the simpler approximations is just as good for a face centered
lattice.  These results demonstrate that exchange energies between
orbitals on different sites are extremely small.  The exchange
interactions are only significant between nearest neighbor Gaussian
orbitals and the exchange energy per electron is given by $E_X = -
(N_n/2)\sqrt{C / \pi}\exp(-CR^2)$, where $N_n$ is the number of
nearest neighbors and $R$ is the nearest neighbor distance.  This
expression gives extremely small energies for the low densities
studied here.

We have also calculated DMC energies using the orbitals obtained from
HF theory~\cite{trail_wigner}, and Jastrow factors optimized using
energy variance minimization. The results are summarized in
Table~\ref{table:hf_opted_orbs_and_dmc_results}. The extrapolation of
the DMC results to infinite system size was carried out using
Ceperley's extrapolation scheme with his value of
$b=1.5$.\cite{ceperley_fermion_plasma} Comparing the data in
Table~\ref{table:hf_opted_orbs_and_dmc_results} with that in
Tables~\ref{table:opt_exponents_and_dmc_energies}
or~\ref{table:zero_tau_dmc_energies}, we see that the DMC energy
obtained at $r_s=100$ with the HF orbitals is significantly higher
than that obtained with DMC optimized orbitals. Some population
control bias is present in the results in
Table~\ref{table:hf_opted_orbs_and_dmc_results}, but we estimate this
to be about 2\% of the difference between the $r_s=100$ DMC energies
obtained with the HF and DMC-optimized orbitals.

\begin{table}[ht]
\begin{tabular}{rrrrr}\hline\hline
$r_s$ & $C$ (HF) & HF energy & \multicolumn{2}{c}{DMC energy} \\ & & &
    (512 electrons) & (Infinite system) \\ \hline

50 & $0.00141$ & $-0.0136768$ & $-0.014060(4)$ & $-0.014052(4)$ \\

100 & $0.0005$ & $-0.0074593$ & $-0.007589(1)$ & $-0.007586(1)$\\
\hline\hline
\end{tabular}
\caption{HF results and DMC results obtained using HF theory
orbitals. All entries are in a.u.
\label{table:hf_opted_orbs_and_dmc_results}}
\end{table}

The strength of correlations in a system may be measured by the ratio
of the correlation energy to the total energy, $E_{\rm c}/E$.  The DMC
results of Zong \textit{et al.}~\cite{zong_2002} indicate that in the
fluid phases $E_{\rm c}/E$ tends to a positive constant as
$r_s\rightarrow\infty$, while for the Wigner crystal our results show
that $E_{\rm c}/E$ tends to zero as $r_s\rightarrow\infty$.  In this
sense one may think of Wigner crystals as being weakly correlated
systems at low densities.

\section{Magnetic behavior of the crystalline phases}

The tiny energy differences between ferromagnetic and
antiferromagnetic crystals proved to be too small to resolve in our
QMC calculations.  It might be possible to resolve them using a
correlated sampling approach within VMC.  Such an approach should
provide an accurate value for the energy difference between two
systems, 1 and 2, if $|\Psi_1|^2 \simeq |\Psi_2|^2$ throughout the
configuration space, which should hold for ferromagnetic and
antiferromagnetic crystalline phases at sufficiently low densities.
HF theory predicts ferromagnetic behavior in the low density limit but
according to the theory of Thouless~\cite{thouless_1965} such a system
should be antiferromagnetic.  In their path integral QMC calculations,
Candido, Bernu, and Ceperley~\cite{bernu} have indeed found
antiferromagnetic behavior for BCC Wigner crystals at low densities,
although the energy differences are much smaller than our statistical
error bars.

\section{Floating Wigner crystals \label{section:floating}}

The Hamiltonian of the uniform electron gas is invariant under the
simultaneous translation or rotation of the electron
positions. However, our trial Wigner crystal wave functions break
these symmetries and, for example, the resulting charge densities are
inhomogeneous, see Fig.~\ref{graph:linedensity_wigner_rs100_ferro}.
These wave functions represent Wigner crystals which are ``pinned'' by
some small external influence.  Pinning of Wigner crystals may arise
from the presence of impurities or boundaries, and therefore the
broken symmetry solutions are physically relevant, but Wigner crystals
may also be mobile, in which case it is better to describe them as
\textit{floating} crystals.

One way of restoring the homogeneous and isotropic nature of the trial
function is to consider a linear combination of displaced and rotated
copies of the fixed wave function $\Psi$. This gives a ``floating''
wave function,
\begin{equation}
\Psi_F = \int \Psi(R(\tilde{\Omega})(\{{\bf r}_i - {\bf
\Delta}\}))\,d\tilde{\Omega}\,d{\bf \Delta},
\label{equation:floating_wf}
\end{equation}
where $R(\tilde{\Omega})$ represents a rotation of all of the electron
positions and the integrals are over all possible solid angles
$\tilde{\Omega}$ and displacements ${\bf \Delta}$. $\Psi_F$ gives rise
to a homogeneous and isotropic charge density $n({\bf r})=N / \Omega$,
where $N$ is the number of electrons and $\Omega$ is the volume of the
system.

Wave functions for floating Wigner crystals in extended systems have
been discussed briefly by, for example, Bishop and
L\"uhrmann~\cite{bishop_1982}, and by Mikhailov and
Ziegler.\cite{mikhailov_2002} Rather more attention has been devoted
to restoring the (rotational) symmetry in two-dimensional models of
quantum dots.\cite{yannouleas_1999,mikhailov_2002} In finite systems
the energy gain per electron from restoring the symmetry can be
substantial but for an infinite system it turns out to be negligible.

Using a trial function where the single particle orbitals in the
Slater determinant are $\phi = e^{-Cr^2}$, we have obtained analytical
results at the variational level for the case when the translational
symmetry is restored. We found that the difference in total energy is
equal to the kinetic energy of the center of mass of the fixed crystal
($3C/2$), making the energy difference per electron negligible (this
result also holds when a translationally invariant Jastrow factor is
included). We also found that the expectation value of any operator
that only depends on relative electron coordinates is the same for
$\Psi$ and $\Psi_F$ at the variational level. We expect the same
conclusions to be true in DMC as well; in particular, we found that
under \emph{open} boundary conditions the nodal surface of the fixed
and the translationally averaged trial functions are identical.

Despite the similarities in their energies, there is an important
qualitative difference between the fixed and floating Wigner crystal
wave functions. The fixed wave function can be written as a sum of
disconnected partial wave functions, in the form $\Psi = \sum_M
\psi_M$, where the overlap between the $\psi_M$ tend exponentially to
zero as $N \to \infty$. From this it follows that it represents an
insulator.\cite{kohn_1964,martin_2000} On the other hand, the same is
not true for the floating wave function, resulting in conducting
behavior.

\section{Pair correlation functions}

\subsection{Definitions}

The spin-dependent PCF is defined as
\begin{equation}
\label{eq:pcf1}
g_{\sigma,\sigma^\prime}({\bf r},{\bf r}^\prime) =
  \frac{\left<\sum_{i,j \neq
  i}\delta_{\sigma,\sigma_i}\delta_{\sigma^\prime,\sigma_j}
  \delta({\bf r}-{\bf r}_i)\delta({\bf r}^\prime-{\bf
  r}_j)\right>}{n_{\sigma}({\bf r}) n_{\sigma^\prime}({\bf r}^\prime)}.
\end{equation}
It would be very costly to evaluate the full six-dimensional function
$g_{\sigma,\sigma^\prime}({\bf r},{\bf r}^\prime)$ for a Wigner
crystal within QMC.  However, for a homogeneous and isotropic system
$g$ depends only on the separation between electrons, $r = |{\bf
r}-{\bf r}^\prime|$, so that
\begin {equation}
\label{eq:paircorrfn}
g_{\sigma,\sigma^\prime}(r) = \frac{\Omega}{4 \pi
r^2}\frac{1}{N_{\sigma}N_{\sigma^\prime}}\left<\sum_{i,j \neq
i}\delta(r-|{\bf r}_i-{\bf r}_j|)\right>,
\end {equation}
where we have used $n_{\sigma}({\bf r}) = N_\sigma / \Omega$.  This
one-dimensional function is much less costly to evaluate accurately
than Eq.~(\ref{eq:pcf1}), hence we can obtain the PCF for a floating
Wigner crystal.  Since the expectation value of any operator that
depends only on the difference between electron coordinates is the
same for fixed and floating crystals, see Sec.~\ref{section:floating},
we can evaluate $g_{\sigma,\sigma^\prime}(r)$ using our fixed trial
function.

Furthermore, the expectation value of the potential energy of the
system can be written in terms of the spin-independent PCF, $g(r) =
\frac{1}{4} \sum_{\sigma,\sigma^\prime} g_{\sigma,\sigma^\prime}(r)$:
\begin{eqnarray}
\label{eq:potential_energy}
\left< \sum_{i,j \neq i} v(|{\bf r}_i - {\bf r}_j|) \right> & = & \int
n({\bf r})n({\bf r}^\prime)g({\bf r},{\bf r}^\prime)v(|{\bf r} - {\bf
r}^\prime|) \,d{\bf r}\,d{\bf r}^\prime \nonumber \\ & = &
\frac{N^2}{\Omega}\int 4 \pi r^2 \, g(r)v(r) \, dr,
\end{eqnarray}
which holds even for an inhomogeneous system such as the fixed Wigner
crystal. Studying the PCF of a floating crystal therefore provides
insight into the physics of the fixed crystal as well.

\subsection{Discussion and Results for the PCFs}

We evaluated Eq.~(\ref{eq:paircorrfn}) within VMC and DMC by
accumulating $g_{\sigma,\sigma^\prime}(r)$ in radial bins.  Our best
estimates of $g$ were obtained using the extrapolated estimator $g(r)
\approx 2 g_{\rm{DMC}} - g_{\rm{VMC}}$.  We calculated
$g_{\sigma,\sigma^\prime}(r)$ at $r_s = 110$ for a ferromagnetic and
an antiferromagnetic Wigner crystal using our optimized trial wave
functions. For comparison, we have also calculated
$g_{\sigma,\sigma^\prime}(r)$ for the unpolarized fluid phase at $r_s
= 110$, using a trial wave function consisting of determinants of
plane waves and a Jastrow factor optimized using energy variance
minimization.

All of the biasing effects that apply to the DMC energy may also
affect the PCFs. We found that this was indeed the case, as we have
obtained results that showed very small but statistically significant
differences when, for example, different timesteps were used or when
different Jastrow factors were used in conjunction with small
population sizes. Unlike the energy, it was not possible to use an
extrapolation scheme to remove the timestep bias, as it showed no
clear pattern. We found that finite size effects in the PCFs were very
small, however, as results obtained for 64 electrons were not
significantly different from those obtained with 512 electrons.

\begin{figure}
\includegraphics{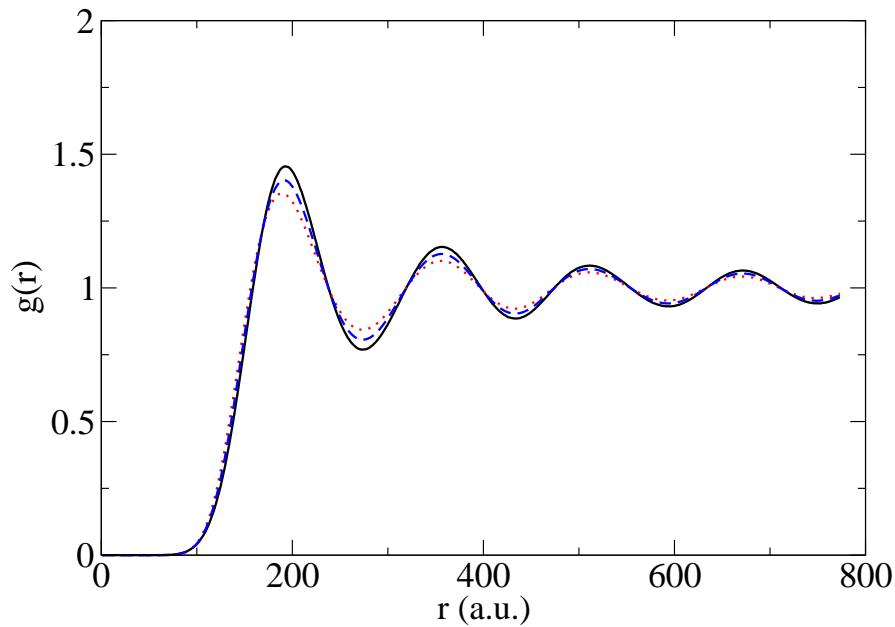}
\caption{Spin-independent PCF in the antiferromagnetic floating
crystal at $r_s = 110$, illustrating the extrapolation
procedure. Solid line: extrapolated estimator; dashed line: DMC
result; dotted line: VMC result.  The statistical errors are less than
0.003.
\label{graph:extrapolate_pcf}}
\end{figure}

A further source of bias, which does not apply to the energy, arises
from the use of the extrapolated
estimator. Fig.~\ref{graph:extrapolate_pcf} shows that the
extrapolation can make a significant difference, and to check the
reliability of this method we have evaluated the PCF using different
quality Jastrow factors. As might be expected, the VMC and DMC results
varied significantly, but the final extrapolated results were almost
independent of the Jastrow factor, the differences being only slightly
larger than the statistical error. This source of bias is therefore
small and on a comparable level to the others.

\begin{figure}[ht]
\includegraphics{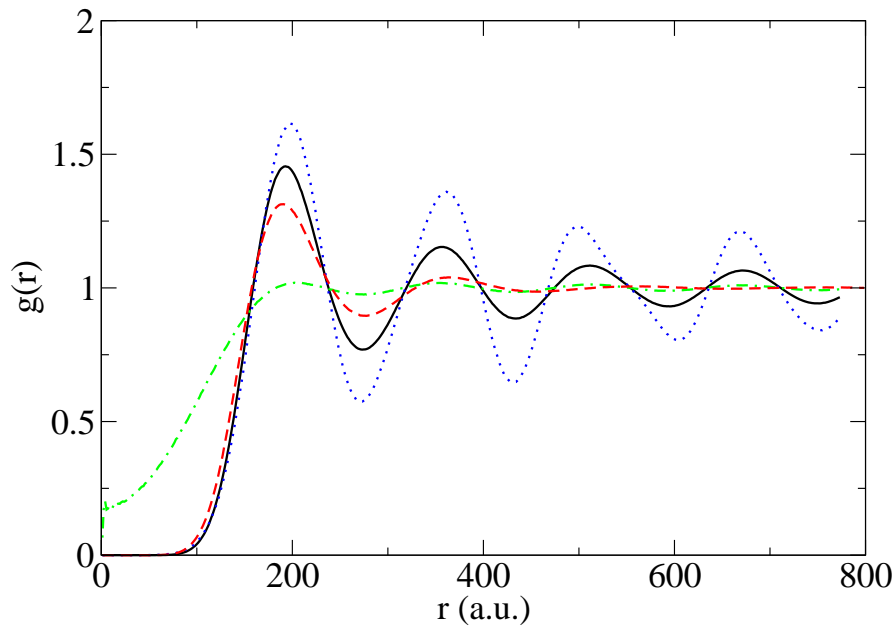}
\caption{Spin-independent PCF for the antiferromagnetic floating
crystal and unpolarized fluid at $r_s = 110$. Solid line: extrapolated
estimator for the crystal; dotted line: HF result; dashed-dotted line:
VMC result obtained using the DMC optimized orbitals but without a
Jastrow factor. The dashed line shows the extrapolated estimator for
the unpolarized fluid. The statistical errors are less than 0.003
except for the dashed-dotted line around $r=0$.
\label{graph:all_elec_pcf}}
\end{figure}

The final PCF calculations were carried out in a 512 (518) electron
system for the Wigner crystal (fluid), using a timestep of 30 a.u.~and
a target population of 960.  Fig.~\ref{graph:all_elec_pcf} shows the
extrapolated spin-independent PCF for the antiferromagnetic crystal
and unpolarized fluid phases at $r_s=110$. For comparison, the HF
result is also shown, together with the result obtained from the
DMC-optimized orbitals, but without the Jastrow factor.  The
spin-independent PCF for the ferromagnetic crystal is not included as
it was found to be almost identical to that of the antiferromagnetic
crystal.  The HF orbitals are very strongly localized and give the
most rapidly varying PCF, whereas the PCF from the more diffuse
DMC-optimized orbitals is much smoother.

It is interesting that the extrapolated PCF of the floating crystal
shows strong oscillations at distances much larger than $r_s$.  This
can be understood in terms of Eq.~(\ref{eq:potential_energy}).  The
potential energies of the fixed and floating crystals are the same,
but their charge densities and PCFs are quite different.  For the
floating crystal $n({\bf r})$ is constant while $g({\bf r},{\bf
r}^\prime)$ oscillates strongly, whereas for a fixed crystal the
charge density $n({\bf r})$ oscillates and $g({\bf r},{\bf r}^\prime)$
is expected to be much smoother.

\begin{figure}[ht]
\includegraphics{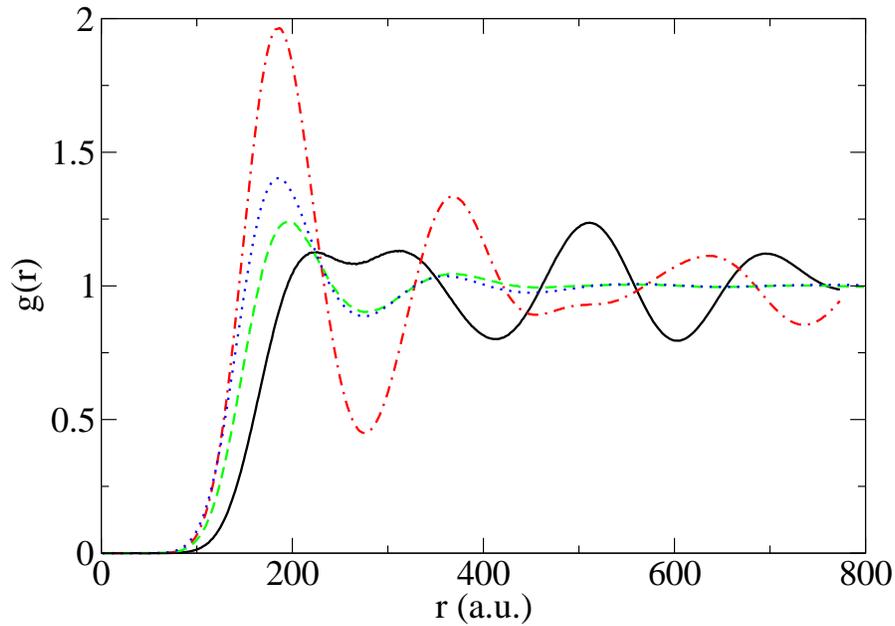}
\caption{Parallel and antiparallel spin PCFs in the antiferromagnetic
crystal and unpolarized fluid at $r_s = 110$. Solid line: parallel
spin crystal PCF; dashed line: parallel spin fluid PCF; dash-dotted
line: antiparallel spin crystal PCF; dotted line: antiparallel spin
fluid PCF. The statistical errors are less than 0.005.
\label{graph:AF_fluid_pcf}}
\end{figure}

Another interesting point is the large difference between the
extrapolated parallel/antiparallel-spin PCFs of the crystal and the
fluid, see Fig.~\ref{graph:AF_fluid_pcf}. For the crystal, the PCF
strongly reflects the underlying crystal structure composed of
alternating spins, whereas for the fluid phase the two are almost
identical. The energies of the two systems are, however, almost the
same.

\section{Conclusions}

We have carried out a careful DMC study of the BCC Wigner crystal in
the density range $100 \leq r_s \leq 150$. We have experimented with
several types of orbital in the trial wave functions but have been
unable to improve upon the Gaussian form used in previous work. We
have, however, optimized the Gaussian exponent by directly minimizing
the DMC energy, which reduces the fixed node errors.  We have also
taken care to eliminate other biases in our DMC simulations,
particularly those from timestep errors, population control bias, and
finite size effects. We estimate that the uniform electron gas
undergoes a transition from a ferromagnetic fluid to a BCC crystal at
$r_s=106 \pm 1$.

We have used Slater-Jastrow-type trial wave functions for our studies.
Multiplying the Slater determinant by a pairwise repulsive Jastrow
factor makes the charge density more inhomogeneous because the
electrons in Wigner crystals are localized in space individually.
This behavior contrasts with that found in many other systems where a
pairwise repulsive Jastrow factor tends to smooth out the charge
density.

The results of HF theory and Hartree theory are very similar because
the exchange interaction between orbitals on different sites is small.
The orbitals obtained within unrestricted HF theory (and Hartree
theory) are very strongly localized and the kinetic energy within HF
theory is larger than in our DMC calculations with the fully optimized
trial wave functions.  We have defined the correlation energy to be
the difference between the exact and unrestricted HF energies.  With
this definition, and in the low density limit, Wigner crystals are
weakly correlated systems.  The inclusion of correlation in a Wigner
crystal wave function beyond the unrestricted HF level results in the
electronic charge density spreading out from the lattice sites.  In
this sense correlation delocalizes the electrons.  Although the
correlation energies of the crystal phases are smaller than those of
the fluid phases at the same density, the use of HF orbitals within
the trial wave functions results in larger fixed node errors for the
crystal phases.

The variational energy for a floating Wigner crystal is lower than
that of the fixed crystal by the kinetic energy of the center of mass,
which is a negligible energy per particle for large systems.  The
expectation value of any operator that depends only on the difference
between electron coordinates is the same for the fixed and floating
crystals. We can therefore obtain the PCFs of a floating Wigner
crystal rather simply from calculations on the fixed crystal.

\section{Acknowledgments}

Financial support was provided by the Engineering and Physical
Sciences Research Council (EPSRC), UK.  MDT thanks the Royal Society
for a Research Fellowship and ZR thanks the Gates Cambridge Trust for
funding a studentship.  Computational facilities were provided by the
High Performance Computing Facility at the University of Cambridge,
and the Computer Services for Academic Research (CSAR) in Manchester,
UK.


\begin{thebibliography}{}
\bibitem{units} We use Hartree atomic units, $\hbar=|e|=m_e=4\pi
\epsilon_0=1$, throughout. The density parameter $r_s$ is defined by
$4\pi r_s^3/3 = \Omega/N$, where $N/\Omega$ is the number of electrons
per unit volume.
\end{thebibliography}
\end{document}